\documentclass[aps,prd,onecolumn,groupedaddress,showpacs,nofootinbib,amssymb]{revtex4}
\usepackage[dvips]{graphicx}
\usepackage{amssymb}
\usepackage{amsmath}
\usepackage{graphicx}
\usepackage{amsfonts}
\usepackage{bm}
\begin{document}

\title{ $f(R,T)$ models applied to Baryogenesis }
\author{E.~H.~Baffou,$^{(a,b)}$\,\footnote{Email:baffouhet@gmail.com}
M.~J.~S.~Houndjo,$^{(a,b)}$\,\footnote{Email: sthoundjo@yahoo.fr}
D.~.A.~Kanfon, $^{(b,c)}$\,\footnote{Email: kanfon@yahoo.fr}
I. G. Salako $^{a,d}$\,\footnote{Email:inessalako@gmail.com} 
}
\affiliation{ 
$^{a}$ \, Institut de Math\'{e}matiques et de Sciences Physiques, 01 BP 613,  Porto-Novo, B\'{e}nin\\
$^{b}$\, Facult\'e des Sciences et Techniques de Natitingou, BP 72, Natitingou, B\'enin \\
$^{c}$\, D\'epartement de Physique, Universit\'e d'Abomey-Calavi, BP 526 Calavi, Benin \\
$^{d}$\, D\'epartement de Physique, Universit\'e Nationale
d’Agriculture , 01 BP 55 Porto-Novo, Benin
}
\begin{abstract}
This paper is devoted to the reproduction of the gravitational baryogenesis epoch in the context of $f(R, T)$ theory of gravity, where $R$ and $T$ are respectively the curvature scalar and the trace of the energy-momentum tensor, respectively. It is assumed a minimal coupling between matter and gravity. In particular we consider the following two models, $f(R,T) = R +\alpha T + \beta T^2$ 
and $f(R,T) = R+ \mu R^2 + \lambda T$, with the assumption that the universe is filled by dark energy and perfect fluid where the baryon to entropy ratio during a radiation domination era is non-zero. We constrain the models with the cosmological gravitational baryogenesis scenario, highlighting the appropriate values ​​of model's parameters compatible with the observation data of the baryon-entropy ratio.

\end{abstract}
 

\maketitle


\def\pp{{\, \mid \hskip -1.5mm =}}
\def\cL{\mathcal{L}}
\def\be{\begin{equation}}
\def\ee{\end{equation}}
\def\bea{\begin{eqnarray}}
\def\eea{\end{eqnarray}}
\def\tr{\mathrm{tr}\, }
\def\nn{\nonumber \\}
\def\e{\mathrm{e}}

\section{Introduction}

Since antiparticles were first predicted and observed (\cite{baf1}), it has been clear that there
exist high degree of matter-antimatter symmetry.
This observation is a stark contradiction to the phenomena of everyday and
cosmological evidence, particularly the fact that our
universe consists of almost entirely matter with little primordial antimatter. The successful of this discovery is verified by the predictions of  
Big-Bang Nucleosynthesis (BBN) (\cite{baf2}), the highly precise measurements of the cosmic microwave background (\cite{baf3}) and the absence of intense
radiation from matter-antimatter annihilation (\cite{baf4}).
The origin of the 
baryon number asymmetry is an open issue of the modern Cosmology and particle
physics. Various baryogenesis scenarios explain how there are more matter than antimatter in this universe (\cite{baf5})-(\cite{baf12}), which might occur during the matter or the radiation eras.
The existence of processes which violate C and CP tells us that there is a fundamental
asymmetry between matter and antimatter {\footnote {One is the charge conjugation symmetry ($C$-symmetry) and the other is the parity symmetry ($P$ -
		symmetry). The combined symmetry of the two is called, $CP$ -symmetry.}}. Thus the possibility
arises of processes which preferentially produce matter rather than antimatter (although
our present theoretical understanding doesn't allow us to deduce this directly from the
observed CP violation). However, even if this is the case, the ratio of particles and antiparticles will be very close
to unity providing they are in equilibrium, as will be the case when the universe was very
hot. Only as it cools and the equilibrium is removed will the tiny asymmetry in the particle
interactions be amplified to an actual asymmetry in number densities. These requirements
to produce matter-antimatter asymmetry, namely, (a) non-conservation of baryon number,
(b) CP violation and (c) non-equilibrium are known as the Sakharov conditions \cite{baf13}.
In order to connect to dark energy, the authors (\cite{baf14})-(\cite{baf15}) have studied a class of models of spontaneous baryo(lepto)genesis by introducing a interaction between the dynamical dark energy scalars and
the ordinary matter. Recently,  Davoudiasl et al. \cite{baf16} have proposed a mechanism for generating the
baryon number asymmetry in thermal equilibrium during the expansion of the Universe by means of a dynamical breaking of CP.
The interaction responsible for CP violation is given by a coupling between the derivative of the Ricci scalar $R$ and the baryon current $J^{\mu}$
of the form
\begin{eqnarray} \label{0}
\frac{1}{M_{*}^2} \int \sqrt{-g} dx^{4} \partial_{\mu}(R) J^{\mu},
\end{eqnarray}
where $M_{*}$  is the cutoff scale characterizing the effective
theory, $g$ and $R$ being respectively, the metric determinant and the Rurvature scalar.
Other scenario to extend this well known theory by using a similar couplaging between the  Ricci scalar
and the baryonic current has been discussed by many authors.
This scenario extends the well known theory that uses a similar coupling between the Ricci scalar
and the baryonic current. In (\cite{baf17}), $f (R)$ theories of gravity are reviewed in the context of the so called gravitational baryogenesis.
Some variant forms of gravitational baryogenesis by using higher order terms containing the partial derivative of the Gauss-Bonnet scalar coupled 
to the baryonic current are discussed in (\cite{baf18})
whereas in (\cite{baf19}), the gravitational baryogenesis scenario, generated by an $f(T)$ theory of gravity where $T$ is 
the torsion scalar are proposed. \par
The purpose of this paper is to investigate the gravitational baryogenesis mechanism in $f(R,T)$ modified theory of gravity, a theory in which matter and geometry are minimally coupled 
and well known as generalization of General Theory of Relativity. This Theory was firstly introduced by the authors of (\cite{baf20}) and several works with
interesting results  have been found in \cite{baf21}-\cite{baf28}. \par
The paper is organized as follows: A brief review in $f(R,T)$ gravity is performed in section ({\bf 2)}. We investigate the essential features of baryogenesis
in $f(R,T)$ gravity by calculating the corresponding baryon to entropy ratio in universe containing a dark energy and the perfect
fluid with constant equation of state parameter in section {(\bf 3)}. Some conclusions
are presented in the last section.
\section{ Brief Review in $f(R,T)$ Gravity }
Let us consider the total action in modified $f(R,T)$ gravity by
\begin{eqnarray}
S =  \int \sqrt{-g} dx^{4} \Big[\frac{1} {16\pi G} f(R,T)+\mathcal{L}_m \Big]\,\,, \label{1}
\end{eqnarray}
where $R$, $T$ are the curvature scalar and the trace of the energy-momentum tensor, respectively, $G$
being the gravitation constant.\\
From the matter Lagrangian density $\mathcal{L}_m$,  we defined the energy-momentum tensor of the matter as
\begin{eqnarray}
T_{\mu\nu}=-\frac{2}{\sqrt{-g}}\frac{\delta\left(\sqrt{-g}\mathcal{L}_m\right)}{\delta g^{\mu\nu}}.\label{2}
\end{eqnarray} 
Varying the action (\ref{1}) with respect to the metric formalism, the field equations are obtained as
\begin{eqnarray}
f_{R}R_{\mu\nu}-\frac{1}{2} g_{\mu\nu}f(R,T)+ \mathcal{D}_{\mu\nu} f_{R}= 8\pi G T_{\mu\nu}-
f_{T}(T_{\mu\nu}+\Theta_{\mu\nu})\,,\label{3}
\end{eqnarray}
where
\begin{eqnarray}
\mathcal{D}_{\mu\nu} = g_{\mu\nu}\Box-\nabla_{\mu}\nabla_{\nu},  
\end{eqnarray}
\begin{eqnarray}
\Theta_{\mu\nu}\equiv g^{\alpha\beta}\frac{\delta T_{\alpha \beta}}{\delta g^{\mu\nu}}=-2T_{\mu\nu}+g_{\mu\nu}\mathcal{L}_m
-2g^{\alpha\beta}\frac{\partial^2 \mathcal{L}_m}{\partial g^{\mu\nu}\partial g^{\alpha \beta}}\label{4}.
\end{eqnarray}
We note that $f_{R}, f_{T}$  are the partial derivatives of $ f(R,T) $ with respect
to  $ R $ and $T$, respectively. The field equations (\ref{3}) are reduced to Einstein field equations when $f(R,T) \equiv R$.\\
Contracting Eq.(\ref{3}) with the tensor metric components $g^{\mu\nu}$, one gets the relation between the Ricci scalar $R$ and the trace $T$ of the energy momentum tensor
\begin{eqnarray}
f_{R} R-2f(R,T) +3\Box f_{R} =8\pi G   T -f_{T}(T+\Theta)\label{5}. 
\end{eqnarray} 
Let us now consider the spatially flat FLRW spacetime
\begin{eqnarray}
ds^{2}= dt^{2}-a(t)^{2}[dx^{2}+ dy^{2}+dz^{2}],\label{6}
\end{eqnarray}
where $a(t)$ is the scale factor and the matter content of the universe is a perfect fluid for which the matter Lagrangian density can be taken
as $\mathcal{L}_m = -p $.
For this, the Eqs.(\ref{3}) and (\ref{5}) become 
\begin{eqnarray}
3H^2= \frac{8\pi G+f_T}{f_R}\rho+ \frac{1}{f_R}\bigg[ \frac{1}{2}(f-Rf_R)-3\dot{R}Hf_{RR}+ p f_T \bigg], \label{7} 
\end{eqnarray}
\begin{eqnarray}
-2\dot{H}-3H^2 =\frac{8\pi G+f_T}{f_R}p + \frac{1}{f_R}\bigg[2H\dot{R}f_{RR}+\ddot{R} f_{RR}+ \dot{R}^{2}f_{RRR}-
\frac{1}{2}(f-Rf_R)-  p f_T \bigg ], \label{8}
\end{eqnarray}
where the dot denotes the derivative with respect to the cosmic time $t$ and $H=\frac{\dot{a}}{a}$, the Hubble parameter.
In the above equations, $\rho$ is the matter density, $p$ the matter pressure and the trace $T= \rho -3p$.
\section{$f(R,T)$ Baryogenesis}
In $f(R,T)$ gravity where we take account a minimal coupling between matter and geometry, we consider a CP-violating interaction term generating
by the baryon asymmetry of the universe  of the form,
\begin{eqnarray} \label{9}
\frac{1}{M_{*}^2} \int \sqrt{-g} dx^{4} \Big( \partial_{\mu}(R+T)\Big) J^{\mu}.
\end{eqnarray}
We define the baryon to entropy ratio as
\begin{eqnarray}\label{10}
\frac{n_B}{s} \simeq -\frac{15 g_b}{4\pi^2 g_{*}} \bigg[ \frac{1}{M_{*}^2\mathcal{T}} \Big(\dot{R}+\dot{T}\Big) \bigg] \arrowvert{\mathcal{T}_D},     
\end{eqnarray}
where $\mathcal{T}_D$ is the decoupling temperature and $n_B$, the baryon number. 
The `` dot '' denote the derivative with respect the cosmic time. We assume in this paper that a thermal equilibrium exists. For this reason we consider that the universe evolves slowly from an equilibrium state to an equilibrium state with the energy being linked to the temperature $\mathcal{T}$ as
\begin{eqnarray} \label{11}
\rho= \frac{\pi^2}{30}g_{*} {\mathcal{T}}^{4}. 
\end{eqnarray}
In (\ref{11}), $g_{*}$ represents the number of the degrees of freedom of the effectively massless particles.  \\
In the context of GR, if we assume that the matter content of the universe as perfect fluid with constant equation of 
state parameter $w=\frac{p}{\rho}$,
the Ricci scalar $R$ and the trace $T= \rho(1-3w) $ of the energy-momentum tensor of matter are related as
\begin{eqnarray}
R= -8\pi G (1-3w)\rho.
\end{eqnarray}
If the universe is filled by the radiation, the baryon number to entropy ratio (\ref{10}) is equal zero in GR.
This results is different zero for the other content of the matter. 
However, a net baryon asymmetry may be generated during the radiation dominated era in $f(R,T)$ theories of gravity.
To do, we focus on attention on two particulars $f(R,T)$ models namely $f(R,T) = R +\alpha T + \beta T^2$ 
and $f(R,T) =R+ \mu R^2 + \lambda T$  to describe how we can recover the baryogenesis epoch with these models.
We calculate the baryon to entropy ratio for each model by considering a universe filled by the dark energy and perfect fluid with constant 
equation of state parameter $ w= \frac{p}{\rho}$ and assuming that the scale factor  evolve as power-law $a(t) = B t^{\gamma}$ where $B$ is a constant parameter.
\subsection{$f(R,T) = R +\alpha T + \beta T^2$ cases}
For the first case, by using the FRW equation $(\ref{7})$ and both the expressions of the scale factor and equation of state parameter that we assumed, we find 
the analytically expression of the energy density as
\begin{eqnarray}\label{12}
\rho = \frac{-\delta t^2 + \sqrt{\delta ^2 t^4 +12 \zeta \gamma ^2 t^2 }}{2 \zeta t^2},
\end{eqnarray}
where $ \delta = 8\pi G +\frac{\alpha}{2}(3-w)$ and $\zeta = \beta (\frac{3}{2}-7w-\frac{3w^2}{2})$. \\
Equaling this expression with (\ref{11}), we obtain the decoupling cosmic time $t_{D}$ expressed in function of the decoupling temperature ${\mathcal{T_D}}$ as
\begin{eqnarray}\label{13}
t_{D}=\frac{30\sqrt{3}\gamma}{\sqrt{30\delta \pi^2 g_{*} {\mathcal{T_D}}^{4} + \zeta \pi^4 g_{*} {\mathcal{T_D}}^{8} }}. 
\end{eqnarray}
Using Eq.(\ref{13}), we arrive at a final expression of the baryon-to-entropy ratio for the present $f(R,T)$ particular model 
\begin{eqnarray}\label{14}
&& \frac{n_B}{s} \simeq -\frac{ g_b \sqrt{g_{*} \pi ^2 {\mathcal{T_D}}^{4}\big(30\delta+\zeta \pi^2 g_{*} {\mathcal{T_D}}^{4} } \big ) }{3600\sqrt{3}\gamma^2 \pi^2 g_{*}M_{*}^2\mathcal{T}_D} \times \cr
&&\left[ \frac{{15 \gamma^3 (3w-1)}} { \sqrt{   \frac { {\gamma^4 {\big(15 \delta+\zeta \pi^2 g_{*} {\mathcal{T_D}}^{4} } \big )}^2 } { {\pi^4 {g_{*}^2} {\mathcal{T_D }^{8}}  { \big(30 \delta+\zeta \pi^2 g_{*} {\mathcal{T_D}}^{4} } \big )}^2 } }} 
+ 2 \pi^2 g_{*} {\mathcal{T_D}}^{4} (2\gamma-1) \big(30\delta+\zeta  \pi^2  g_{*} {\mathcal{T_D}}^{4}  \big )   \right].
\end{eqnarray}
In the  radiation dominated phase, $ \delta = 8\pi G +\frac{4}{3}\alpha$, $\zeta=-\beta$. Hence, Eq.(\ref{14}) reduces to
\begin{eqnarray}\label{14bis}
\frac{n_B}{s} \simeq - \frac{ g_{b} {\mathcal{T_D}}^{5} (2\gamma-1) \pi \sqrt{g_{*}  {\big(30\delta - \beta g_{*}\pi^2 {\mathcal{T_D}}^{4} \big )}^{3}}} {3600\sqrt{3}\gamma^2 M_{*}^2 }. 
\end{eqnarray}
We can show from Eq.(\ref{14bis}) that the resulting baryon
to entropy ratio is non-zero in contrast for the GR if  $\gamma \neq \frac{1}{2}$. Within the choice of the free parameters and depending on matter content, we can adjust the baryon to entropy ratio  Eq.(\ref{14bis}) to satisfy 
the observational constraints.
For illustrating we assume that the cutoff scale $M_{*}$ takes the value $M_{*} = 10^{12} GeV$, also that the critical temperature is equal to
${\mathcal{T_D}} = M_I = 2.10^{16}GeV $, with $M_I$ being the upper bound for tensor mode fluctuations constraints on the inflationary scale,
$g_b \simeq  \mathcal{O}(1)$ and $g_{*} \simeq 106$, which is the total
number of the effectively massless particle in the Universe \cite{baf29}.
In table \ref{Tableau 1}, we present some values of baryon to entropy ratio $\frac{n_{B}}{s}$ for $\gamma=0.4$, $\alpha = 10^{-20}$, $M_{*} = 10^{12} GeV$,   
${\mathcal{T_D}} = 2.10^{16}GeV $ by using Various values of the parameter $\beta$.

\begin{figure}[h]
	\centering
	\begin{tabular}{rl}
		\includegraphics[width=6cm, height=6cm]{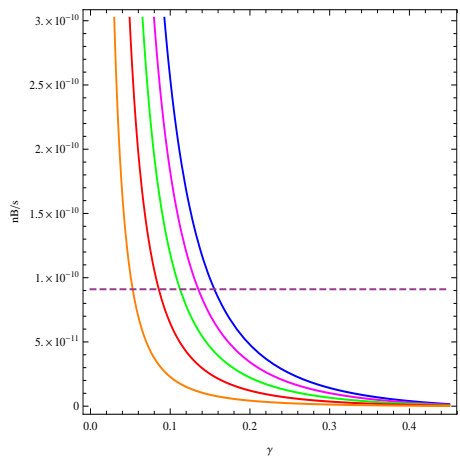}
	\end{tabular}
	\caption{The baryon to entropy ratio for the model $f(R,T) = R +\alpha T + \beta T^2$. The graphs are plotted for $\alpha=10^{-20}$ and for 
		the Varying values of $\beta$,  $\beta = -5.10^{-10}$(Blue), $\beta =-4.10^{-10}$ (Magenta), $\beta =-3.10^{-10}$(Green),
		$\beta =-2.10^{-10}$(Red), $\beta =-10^{-10}$(Orange). 
		The dashed curve represents the observational value.}
	\label{fig1}
\end{figure}

\begin{table}[h]
	\caption{Some values of baryon to entropy ratio for $\gamma=0.4$ and $\alpha = 10^{-20}$}
	\label{tab:xx}
	\begin{center}
		\begin{tabular}{||l||c||c||c||c||c||c||c||}
			\hline
			$\beta$ & $-10^{-9}$ & $-2.10^{-9}$ & $-3.10^{-9}$ & $-4.10^{-9}$ & $-5.10^{-9}$ & $-6.10^{-9}$ \\
			\hline
			
			$\frac{ n_{B}}{s}$ &  $ 1,12.10^{-11}$ &  $3,18.10^{-11}$ &  $ 5,85.10^{-11} $ & $ 9,01.10^{-11}$ & $1,25.10^{-10}$& $1,66.10^{-10}$ \\
			\hline
			
		\end{tabular}
	\end{center}
	\label{Tableau 1}
\end{table}
According to the results of this table, we observe that for $\beta = -4.10^{-9}$,  $n_{B}/s$ $= 9,01.10^{-11}$, which is  very  agreement 
with observations and practically equal to the observed value ( $n_{B}/s \simeq 9,42.10^{-11}$)  whereas when $\beta > -4.10^{-9}$, we denote a significantly
small values. \\
In addition, we plot in fig \ref{fig1},  the $\gamma$- dependence of
the baryon-to-entropy ratio for $M_{*} = 10^{12} GeV$,   
${\mathcal{T_D}} = 2.10^{16}GeV $.
From the curves of the \ref{fig1}, we note that the intersection of each curve with the curve traducing the observational value ( dashed curve) 
are in good agreement value of baryon to entropy ratio for specific value of $\gamma$ including $0$ and $0.5$.
Also, we note that each curve goes towards $0$ when the parameter $\gamma$ tends to $0.5$ which is compatible with theoretical results.

\subsection{$f(R,T) = R+ \mu R^2 + \lambda T$ cases}
\begin{figure}[h]
	\centering
	\begin{tabular}{rl}
		\includegraphics[width=6cm, height=6cm]{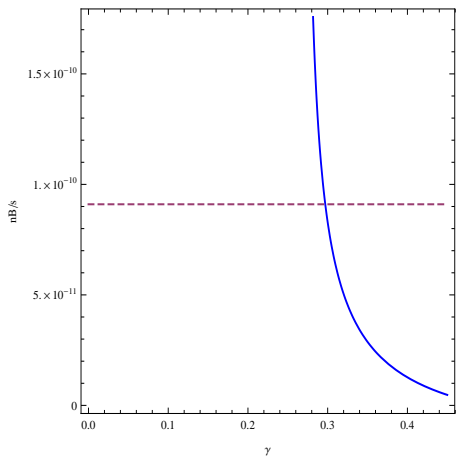}
	\end{tabular}
	\caption{The $\gamma$- dependence of baryon to entropy ratio for the model $f(R,T) = R+ \mu R^2 + \lambda T$ for
		$M_{*} = 10^{12} GeV$,  ${\mathcal{T_D}} = 2.10^{16}GeV $. 
		The dashed curve represents the observational value of baryon to entropy ratio whereas the blue curve represents the evolution of baryon to entropy
		ratio for $ \mu = \lambda =10^{-5}$. }
	\label{fig2}
\end{figure}

The first FRW Eq. (\ref{7}) becomes
\begin{eqnarray}\label{15}
3H^2 (1+2\mu R) = \rho\big( 8\pi G +\frac{1}{2}\lambda(3-w) \big) -\frac{1}{2}\mu R^2 -6\mu \dot{R} H.  
\end{eqnarray}
Making use the scale factor that we assumed, we can solve (\ref{15}) explicitly
\begin{eqnarray}\label{16}
\rho = \frac{ 3 t^2 \gamma^2-\Gamma}{\Delta t^4}, 
\end{eqnarray}
where $\Gamma = 4\mu \gamma^2 \big(\gamma^2-4\gamma+1)$ and $ \Delta= 8\pi G +\frac{1}{2}\lambda (3-w)$. \\
For the special model considered, one can express the decoupling cosmic time $t_{D}$ as
\begin{eqnarray} \label{16}
t_{D} =  \sqrt{ \frac{45 \gamma^2 +\sqrt{2025\gamma^4-30g_{*}\Gamma \Delta \pi^2 {\mathcal{T_D}}^{4}}}{g_{*} \Delta \pi^2 {\mathcal{T_D}}^{4}}}, 
\end{eqnarray}
where we assumed that 
$2025\gamma^4-30g_{*}\Gamma \Delta \pi^2 {\mathcal{T_D}}^{4}> 0 $.  
Then, we reformulate the  baryon to entropy ratio (\ref{10}) as
\begin{eqnarray} \label{17}
\frac{n_B}{s} \simeq - \frac{15 g_b  {(g_{*} {\Delta})}^{\frac{3}{2}} {\mathcal{T_D}}^{9} \pi^3}
{ M_{*}^2 {{\Bigg [45\gamma^2 + {{ \Big( 2025\gamma^4-30g_{*} \Gamma \Delta \pi^2 {\mathcal{T_D}}^{4} }} \Big)}^{\frac{1}{2}} \Bigg] }^{\frac{5}{2}}} \times \nonumber \\ 
\left[ \Gamma(2-6w) +
\frac{3\gamma \Big(-2\Delta+\gamma(-1+3w+4 \Delta) \Big)\Big(45\gamma^2 + {\big(2025\gamma^4-30g_{*}\Gamma \Delta \pi^2 {\mathcal{T_D}}^{4} \big)}^{\frac{1}{2}}\Big)}{g_{*}\Delta \pi^2 {\mathcal{T_D}}^{4}} \right].
\end{eqnarray}
In the radiation dominated epoch, $\Delta = 8\pi G +\frac{4\lambda}{3}$ and the  baryon to entropy ratio (\ref{17}) becomes
\begin{eqnarray} \label{18}
\frac{n_B}{s} \simeq - \frac{90}{M_{*}^2} \gamma g_{b} \pi \Delta^{\frac{3}{2}} {g_{*}}^{\frac{1}{2}} {\mathcal{T_D}}^{5}(2\gamma-1){{\Bigg [45\gamma^2 + {{ \Big( 2025\gamma^4-30g_{*} \Gamma \Delta \pi^2 {\mathcal{T_D}}^{4} }} \Big)}^{\frac{1}{2}} \Bigg] }^{-\frac{3}{2}}.
\end{eqnarray}
We see from this results that the baryon to entropy ratio is non-zero in case where $\gamma \neq \frac{1}{2}$ for the special $f(R,T)$ model considered.
Notice that for $\gamma = 0.3$, and $\mu = \lambda = 10^{-5}$, the baryon to entropy ratio $n_{B}/s = 8,28.10^{-11}$, which is a compatible with
the observational value.
In the same way, we plot in figure (\ref{fig2}) the evolution of  baryon to entropy ratio (\ref{18}) versus $\gamma$ for $M_{*} = 10^{12} GeV$,   
${\mathcal{T_D}} = 2.10^{16}GeV $ in comparison with the curve traducing the observational value.

\section{Conclusion}
The paper is devoted to the study of gravitational baryogenesis mechanism in the context of $f(R,T)$ theories. According that
the CP-violating interaction that will generate the baryon asymmetry of the
Universe and considering that  the matter content of the universe as perfect fluid with constant
equation of state parameter, we evaluate the baryon to entropy ratio for two particulars $f(R,T)$ models. In contrast with GR, we show for both models
that the baryon to entropy ratio is non-zero in radiation dominated epoch if the parameter $\gamma \neq \frac{1}{2}$ and the baryogenesis epoch
can be reproduced in such theory.



\end{document}